\documentclass{article}

\usepackage{graphicx}
\usepackage[round]{natbib}  
\usepackage{soul}
\usepackage{color}
\usepackage{ccaption}
\usepackage{pifont}
\usepackage{mathptmx}
\usepackage{amsmath, amsfonts, amssymb, mathrsfs}
\usepackage{graphicx}
\DeclareMathOperator{\logit}{logit}

\usepackage{multirow}
\usepackage{multicol}
\usepackage[framemethod=tikz]{mdframed}
\usepackage{xcolor}
\usepackage{booktabs}
\usepackage{subcaption}

\title{Cost Effectiveness Analyses for Sequential Multiple Assignment Randomized Trials}
\author{Lina M. Montoya \thanks{School of Data Science and Society and Department of Biostatistics, University of North Carolina at Chapel Hill} 
\and Elvin Geng \thanks{Division of Infectious Diseases, Washington University in St. Louis School of Medicine} 
\and Harriet F. Adhiambo \thanks{Department of Child, Family, and Population Health Nursing, University of Washington} 
\and Eliud Akama \thanks{Department of Global Health, University of Washington} 
\and Starley B. Shade \thanks{Department of Epidemiology and Biostatistics, University of California, San Francisco} 
\and Assurah Elly \thanks{Kenya Medical Research Institute} 
\and Thomas Odeny \thanks{Department of Medicine, University of Missouri-Kansas City} 
\and Maya Petersen \thanks{University of California, Berkeley}}
\date{}

\begin{document}
\maketitle

\begin{abstract}
Sequential multiple assignment randomized trials (SMARTs) have grown in popularity in recent years, and many of their study protocols propose conducting a cost effectiveness analysis of the adaptive strategies embedded within them. The cost effectiveness of these regimes is often proposed to be assessed using incremental cost effectiveness ratios (ICERs). In this paper, we present an estimation and inference procedure for such cost effectiveness measures for the embedded dynamic treatment regimes within a SMART design. In particular, we describe a targeted maximum likelihood estimator for the ICER of a SMART's embedded regimes with influence curve-based inference. We illustrate the performance of these methods using simulations. Throughout, we use as illustration a cost effectiveness analysis for the Adaptive Strategies for Preventing and Treating Lapses of Retention in HIV Care (ADAPT-R; NCT02338739) trial, presenting estimated ICERs (with inference) for embedded regimes aimed at increasing HIV care adherence. This manuscript is one of the first to present cost effectiveness analysis results from a SMART.
\end{abstract}

\section{Introduction}
Sequential Multiple Assignment Randomized Trials (SMARTs) have grown in popularity in recent years. The appeal is warranted -- by sequentially assigning treatment based on accrued participant information, SMARTs allow investigators to evaluate the effects of both point-treatment and longitudinal, as well as static and dynamic, interventions on clinical outcomes of interest (see e.g., \cite{kidwell2023sequential, almirall2014introduction} for an introduction to SMARTs). In particular, SMARTs allow for straightforward evaluation of so-called ``embedded regimes," or dynamic treatment regimes that are a function of the participant information that define the SMART's randomization scheme. 

As an illustrative example, in the SMART called Adaptive Strategies for Preventing and Treating Lapses of Retention in HIV Care (ADAPT-R) trial (NCT02338739; \cite{geng2023adaptive}), adult patients living with HIV in rural Kenya were initially randomized to interventions intended to prevent lapses in HIV care (short message service [SMS] messages, conditional cash transfers [CCTs], or standard-of-care [SOC] counseling). If patients had a lapse in HIV care, they were re-randomized to a more intensive intervention (SMS and CCTs, peer navigator, or SOC outreach), intended to re-engage them back into care. If patients succeeded in their initial care and were initially randomized to an active arm, they were re-randomized to either continue or discontinue the initial intervention. Those who succeeded and were initially given SOC remained in SOC. Under this design, it is straightforward to identify the effects of 15 embedded regimes on HIV care retention (listed in Table \ref{table0}), in addition to contrasts comparing strategy pairs head-to-head. Results of this study showed that the best strategies for improving HIV care retention involve first giving patients active, preventative interventions (i.e., SMS messages or CCTs); then, if patients have a lapse in care, replace the initial intervention with a peer navigator, otherwise, if patients remain in care, maintain the initial intervention. On the other hand, strategies that involve discontinuation of CCTs for those who successfully remain in care tend to compromise retention.

\begin{center}
\begin{table}[]{
\begin{tabular}{|l|l|l|l|}
\hline
\textbf{Embedded Regime ($\tilde{d}$) } & \textbf{Stage 1} & \textbf{Stage 2  if Lapse} & \textbf{Stage 2 if No Lapse} \\ \hline
1 & SOC & SOC outreach & Continue \\ \hline
2 & SMS & SOC outreach & Continue \\ \hline
3 & CCT & SOC outreach & Continue \\ \hline
4 & SOC & SMS + CCT & Continue \\ \hline
5 & SMS & SMS + CCT & Continue \\ \hline
6 & CCT & SMS + CCT & Continue \\ \hline
7 & SOC & Navigator & Continue \\ \hline
8 & SMS & Navigator & Continue \\ \hline
9 & CCT & Navigator & Continue \\ \hline
10 & SMS & SOC outreach & Discontinue \\ \hline
11 & CCT & SOC outreach & Discontinue \\ \hline
12 & SMS & SMS + CCT & Discontinue \\ \hline
13 & CCT & SMS + CCT & Discontinue \\ \hline
14 & SMS & Navigator & Discontinue \\ \hline
15 & CCT & Navigator & Discontinue \\ \hline
\end{tabular}
\caption{List of 15 dynamic treatment regimes embedded within the Adaptive Strategies for Preventing and Treating Lapses of Retention in HIV Care (ADAPT-R) study (i.e., ADAPT-R's 15 embedded regimes). Acronyms: SOC is standard-of-care; SMS is Short Message Service; CCT is conditional cash transfer.}
\label{table0}
}
\end{table}
\end{center}

While these insights are helpful to understand the clinical effectiveness of such strategies under unlimited resources, for understanding the potential scalability of these adaptive interventions, especially in resource-limited settings, it is crucial to examine whether these strategies are effective relative to their monetary cost. This is especially true considering that the most effective strategies found in ADAPT-R included peer-navigators and/or sustained conditional cash transfer interventions, which are relatively resource-intensive \citep{o2012treatment, chang2010effect, decroo2012expert}. For these reasons, a secondary objective of the ADAPT-R study was to assess the cost effectiveness of the sequential strategies for helping patients with HIV remain in care.

Indeed, many SMART protocols propose cost effectiveness analyses (see, for example, \cite{belzer2018adaptive, van2023sequential, levy2019implementation} in recent years). Of the protocols that describe their cost effectiveness analysis plan, many propose estimating the incremental cost effectiveness ratio (ICER; e.g., \cite{quanbeck2020balanced, abuogi2023adapt, zhou2020adaptive}), a measure of the monetary worth of an intervention relative to its clinical effectiveness (specifically, the ICER's numerator is the expected counterfactual cost difference between the two intervention strategies and the denominator is the expected counterfactual outcome difference under the same two strategies compared in the numerator) \citep{gold1996cost}. Further, of the protocols that describe an estimation strategy for approximating the ICER, an inverse probability of treatment weighting (IPW)-type estimator is typically proposed (e.g., \cite{pfammatter2019smart}), also known as the ``weight and replicate" method \citep{nahum2012experimental, almirall2014introduction}, with inference based on the non-parametric bootstrap (e.g., \cite{buchholz2020study, patrick2020sequential, johnson2018protocol}). To our knowledge, only one research group has actually presented cost effectiveness results using data generated from a SMART \citep{li2023cost}. Instead of estimating the ICER, they estimated a similar parameter -- the incremental net monetary benefit, which requires a known or assumed willingness-to-pay threshold.  

Methodological papers detailing estimation and inference procedures for cost effectiveness analyses using data from a SMART design are also lacking. Further, semiparametric efficient estimators, such as targeted likelihood likelihood estimators (TMLE; \cite{van2006targeted}), provide a robust approach to improving precision of primary analyses in trials \citep{moore2009covariate}, including in SMARTs \citep{montoya2023efficient}, that increase estimator precision via adjustment of baseline and time-varying characteristics and incorporation of machine learning. Thus, in this paper, using as illustration the cost effectiveness analysis for the ADAPT-R trial, we present and describe a TMLE for the ICER of a SMART's embedded regimes. Additionally, because inference on the TMLE for contrasts between two embedded regimes is based on the efficient influence curve, we derive the efficient influence curve for the ICER using the functional delta method. In this way, it is possible to obtain inference for the ICER based on an alternative to the bootstrap, which may be computationally expensive. We evaluate the performance of the presented estimators and inference procedures using simulation studies. Finally, by applying these methods to data generated from the ADAPT-R trial, we evaluate whether the sequential strategies identified as effective by the primary analysis are indeed cost effective -- in terms of monetary cost. Costing details, code, simulations, and results for this manuscript can be found at https://github.com/lmmontoya/costeff-SMARTs.

The article is organized as follows: in Section 2, we describe the ADAPT-R trial, cost effectiveness questions of interest, and costing data used for the cost effectiveness analysis. In Section 3, we present the causal and statistical cost effectiveness parameters of interest that aim to answer the cost effectiveness questions described in Section 2. In Section 4, we discuss estimation and inference of such parameters. In Section 5, we present simulations illustrating performance of the estimators described. In Section 6, we apply these methods to the ADAPT-R study and present results. We close with a discussion.

\section{The ADAPT-R Trial and Scientific Questions}

\begin{figure}[h]
    \centering
    \includegraphics[scale = .4]{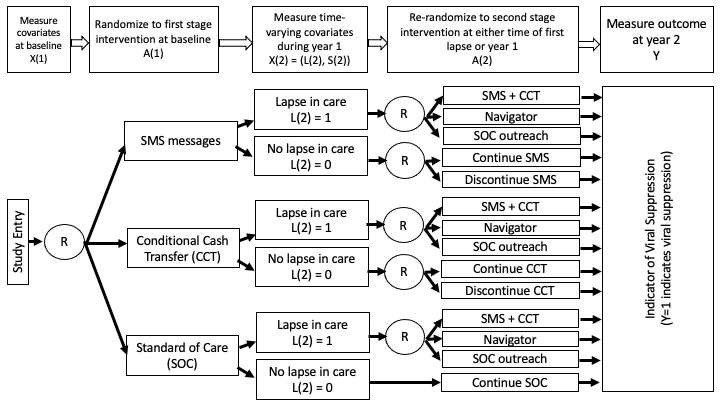}
    \caption{The Adaptive Strategies for Preventing and Treating Lapses of Retention in HIV Care (ADAPT-R) study design, a Sequential Multiple Assignment Randomized Trial (SMART). The circles with an ``R" denote points of randomization.}
    \label{adapt_fig}
\end{figure}

The primary analysis for ADAPT-R examined the effect of 15 embedded regimes on viral suppression, a measure of HIV treatment success \citep{geng2023adaptive}. Specifically, viral suppression was defined as HIV RNA $\leq$ 1000 copies/ml 2 years after enrollment. Results showed that, among of the set of embedded regimes, the best (i.e., the embedded regime with the highest point estimate of the probability of viral suppression) strategy said: ``give CCTs to all. If there is a lapse in care, replace CCTs with a peer navigator; otherwise, continue CCTs." The estimated probability of viral suppression for this strategy was 83.03\% (CI: 76.99-89.07\%) and further, compared to standard of care throughout, this strategy was significantly more effective at increasing viral suppression (risk difference [RD]: 8.19\%; CI: 2.20-14.18\%). In fact, the top three embedded regimes (all significantly more effective than the care standard throughout) involved implementation of a peer navigator and/or continuation of the patient's initial active intervention (i.e., CCT or SMS).

The aforementioned results provide insight on the individualized strategies that are most beneficial for improving HIV care retention in the particular context of the ADAPT-R trial, in which there were sufficient resources to provide these interventions to patients over the course of the two years they are enrolled in the study. However, these interventions were designed to be implemented on and scaled-up for the population from which this sample was drawn. Thus, although it is the case that the sequential interventions that were most effective in ADAPT-R involved peer navigators, which are known to be costly \citep{o2012treatment, chang2010effect, decroo2012expert}, and/or continuing initial, active interventions (such as CCTs), these strategies may not necessarily be the most monetarily feasible to implement or provide the most efficient allocation of resources in rural Kenya. Therefore, in this paper, we aim to ask: a) What are the expected costs of these sequential interventions compared to the care standard? b) What are the added costs of each regime (found in question ``a") relative to their effects compared to the care standard? and c) How does each embedded regime's cost/effectiveness ratio (found in question ``b") compare to other regimes embedded within ADAPT-R?

\subsection{Cost Measures}

Individual-level cost measures for each intervention during the two-year observation period were derived using a combination of primary data from the ADAPT-trial, interviews with ADAPT-R personnel and micro-costing surveys. A detailed breakdown of the cost data sources and definitions is available on the corresponding GitHub page (https://github.com/lmmontoya/costeff-SMARTs). Briefly, CCT cost was defined as the number of cash vouchers each person received in the observation period multiplied by the cost per voucher. Peer navigator costs were obtained by multiplying the number of times a navigator attempted to contact each person during the time he/she/they was in the study by the average cost per navigator attempt -- for both phone and in-person contact attempts. Finally, the per-person SMS cost was defined as the following sum: the cost of sending SMS messages to each person plus the cost of calls to each person and the estimated system cost for each person. We note that a more comprehensive cost effectiveness analysis, encompassing overall cost data -- such as personnel effort, management, and implementation -- is forthcoming. The results presented with these data are intended to illustrate the proposed statistical method and may differ from the final cost effectiveness analysis.

\section{Cost Effectiveness Parameters of Interest}

\subsection{Data \& Causal Models}

In general, the following data are generated from a $K$-stage SMART, where $K$ corresponds to the number of randomization stages: for a time $t$, 1) categorical interventions $A(t)$; 2) covariates $X(t)$, which include baseline covariates and time-varying covariates between interventions at time $t-1$ and $t$; 3) an outcome $Y$; and 4) a cost $C$. Overbars are used to denote a variable's past history, e.g. $\bar{A}(t) = (A(1), \ldots, A(t))$ and $A(0) = X(0) = \emptyset$.

A decision rule $d_t$ is a function that takes as input the information accrued on a participant up to time $t$ and outputs a treatment decision from the set of possible treatment levels to which a participant could be assigned. Additionally, let $\bar{d}_t$ denote a regime sequence until time $t$, and $d = \bar{d}_K = (d_1, \ldots, d_K)$ denotes the entire regime sequence. The subset of observed variables that are used to assign treatment at time $t$ and are thus the allowable inputs to the SMART's embedded regimes are denoted $\bar{Z}(t) \subseteq (\bar{A}(t-1), \bar{X}(t))$. 

With this, the following structural causal model (SCM, denoted $\mathcal{M}^F$) describes the process that gives rise to data generated from a SMART design \citep{pearl2000}, where the random variables in $\mathcal{M}^F$ follow the joint distribution $P_{U,X} \in \mathcal{M}^F$:
\begin{align*}
    X(t) &= f_{X(t)}(U_{X(t)}, \bar{X}(t-1), \bar{A}(t-1)) \\
    A(t) &= f_{A(t)}(U_{A(t)}, \bar{Z}(t)) \\
    C &= f_{C}(U_{C}, \bar{X}(K), \bar{A}(K))\\
    Y &= f_{Y}(U_{Y}, \bar{X}(K), \bar{A}(K)),
\end{align*}
for $t = 1, ..., K$. Here, $U = (U_{X(t)}, U_{A(t)}, U_C, U_Y)$ represents the unmeasured random input to the data generating system. In a SMART, the functions $f_{A(t)}$ are known for all $t$, as they are the randomization scheme used in the SMART. Further, $U_{A(t)}$ is known by design in a SMART to be independent of all other unobserved error terms.

Counterfactual cost and effectiveness outcomes are generated by intervening on the above structural equations. Let $\tilde{d}$ be a regime $d$ that is only a function of the tailoring variables that define the SMART, i.e., $\bar{Z}(K)$. Then, for $t = 1, ..., K$, counterfactual outcomes are generated by:
\begin{align*}
    X(t) &= f_{X(t)}(U_{X(t)}, \bar{X}(t-1), \bar{A}(t-1)) \\
    A(t) &= d_t(\bar{Z}(t)) \\
    C_{\tilde{d}} &= f_{C}(U_{C}, \bar{X}(K), \bar{A}(K))\\
    Y_{\tilde{d}} &= f_{Y}(U_{Y}, \bar{X}(K), \bar{A}(K)).
\end{align*}

In ADAPT-R, $K = 2$. The data consist of baseline covariates $X(1)$ and time-varying covariates $X(2) = (L(2),S(2))$, including whether there was a lapse in care in the first year $L(2)$ and other time-varying covariates $S(2)$ (for a detailed list, see \cite{montoya2023efficient}). First-line treatment is $A(1)$, which is either SMS messages, CCTs, or SOC. Second-line treatment is $A(2)$, which is either SMS messages and CCTs, peer navigators, SOC, a continuation of first-line treatment, or a discontinuation of first-line treatment. Additionally, $Z(1) = \emptyset$ and $Z(2) = (A(1), L(2))$. The outcome $Y$ is 2-year viral suppression and $C$ is the 2-year intervention cost; $C_{\tilde{d}}$ and $Y_{\tilde{d}}$ are the cost and viral suppression outcomes if, possibly counter to fact, an individual had been given embedded regime $\tilde{d}$ of the list in Table \ref{table0}.

\subsection{Causal Cost Effectiveness Parameters}\label{causalparams}

Our target causal parameters, or summary measures of the post-intervention distribution contained within the SCM, mirror the scientific questions asked in the previous section. Denote $\tilde{d}_0$ as the care standard embedded regime, i.e., the embedded regime that says: ``Give everyone SOC counseling. If there is a lapse in care, replace with SOC outreach; otherwise, continue SOC counseling." First, we quantify the incremental cost as the difference in expected costs had everyone received a given embedded regime ($\tilde{d}$) compared to the care standard ($\tilde{d}_0$):
\begin{align*}
    \psi^F_{\text{RD cost}} = \mathbb{E}_{P_{U,X}}[C_{\tilde{d}} - C_{\tilde{d}_0}].
\end{align*}
Analogously, we quantify the incremental effect as the percent difference in the expected health outcome under regime $\tilde{d}$ versus standard of care: $\psi^F_{\text{RD eff.}} = 100\times\mathbb{E}_{P_{U,X}}[Y_{\tilde{d}} - Y_{\tilde{d}_0}]$, where ``eff." stands for ``effectiveness." 

However, not only are we interested in the added costs of a given regime compared to the care standard, but we are also interested in those costs relative to the clinical effects (that is, relative to the effects of the regime on viral suppression). Thus, one causal parameter corresponding this cost effectiveness measure is the following:
\begin{align*}
     \psi^F_{\text{ICER}} = \frac{\psi^F_{\text{RD cost}}}{\psi^F_{\text{RD eff.}}} = \frac{\mathbb{E}_{P_{U,X}}[C_{\tilde{d}} - C_{\tilde{d}_0}]}{100\times\mathbb{E}_{P_{U,X}}[Y_{\tilde{d}} - Y_{\tilde{d}_0}]},
\end{align*}
This causal parameter is commonly known in the literature as the incremental cost effectiveness ratio, or ICER \citep{gold1996cost}, a measure of a given intervention's cost effectiveness.

Finally, we are also interested in contrasting pairs of embedded regimes' cost effectiveness. For example, to compare the cost effectiveness of the last two regimes in Table \ref{table0}, let $\tilde{d}^{(14)}$ be regime \#14 from Table \ref{table0} (SMS, then Navigator if lapse and Discontinue if no lapse), and $\tilde{d}^{(15)}$ be regime \#15 (CCT, then Navigator if lapse and Discontinue if no lapse). Then the causal parameter that contrasts the cost effectiveness between these two strategies is:
\begin{align*}
     \psi^F_{\text{contrast}} & = \psi^F_{\text{ICER}_{\tilde{d}^{(14)}}} - \psi^F_{\text{ICER}_{\tilde{d}^{(15)}}} \\
     & = \frac{\mathbb{E}_{P_{U,X}}[C_{\tilde{d}^{(14)}} - C_{\tilde{d}_0}]}{100\times\mathbb{E}_{P_{U,X}}[Y_{\tilde{d}^{(14)}} - Y_{\tilde{d}_0}]} - \frac{\mathbb{E}_{P_{U,X}}[C_{\tilde{d}^{(15)}} - C_{\tilde{d}_0}]}{100\times\mathbb{E}_{P_{U,X}}[Y_{\tilde{d}^{(15)}} - Y_{\tilde{d}_0}]}.
\end{align*}

\subsection{Statistical Cost Effectiveness Parameters}

The observed data are $O \equiv (\bar{X}(K), \bar{A}(K), Y)$ (with distribution $P_0$ in a statistical model implied by the causal model). The sequential randomization and positivity conditions are sufficient for determining that the causal parameters can be written as statistical parameters. A SMART, by design, ensures that both conditions are met. We refer the reader to \citep{montoya2023efficient} for a discussion. 

The statistical parameter corresponding to the monetary expected value of one embedded regime, i.e., $\mathbb{E}_{P_{U,X}}[C_{\tilde{d}}]$, can be expressed as the following G-computation formula \citep{robins1986new}:
\begin{align*}
\psi_{\text{cost}, \tilde{d}}=& \sum_{x(1),\ldots, x(K)}\mathbb{E}_{0}\left[C|\bar{X}(K) = \bar{x}(K), \bar{A}(K) = \tilde{\bar{d}}_K(\bar{Z}(K))\right] \nonumber \\ 
    & \times \prod_{t=1}^K P_{0}\left(X(t) = x(t) | \bar{X}(t-1) = \bar{x}(t-1), \bar{A}(t-1) = \tilde{\bar{d}}_{t-1}(\bar{Z}(t-1))\right), 
\end{align*} where the summation can be generalized to an integral for continuous $X(t)$. The monetary expected value under the embedded regime in which all receive SOC throughout, called $\psi_{\text{cost}, \tilde{d}_0}$, replaces $\tilde{d}$ with $\tilde{d}_0$ in the equation for $\psi_{\text{cost}, \tilde{d}}$. Thus, the corresponding statistical parameter for the RD with respect to cost, i.e., $\psi_{\text{RD cost}}$, is:
\begin{align*}
    \psi_{\text{RD cost}} = \psi_{\text{cost}, \tilde{d}} - \psi_{\text{cost}, \tilde{d}_0};
\end{align*}
similarly, by switching $C$ to $Y$ and multiplying by 100 in the above G-computation formula, the risk difference with respect to the clinical outcome, viral suppression, is $\psi_{\text{RD eff.}} = 100\times(\psi_{\text{eff.}, \tilde{d}} - \psi_{\text{eff.}, \tilde{d}_0})$. 

We can additionally express the causal ICER as a statistical parameter:
\begin{align*}
     \psi_{\text{ICER}} = \frac{\psi_{\text{RD cost}}}{\psi_{\text{RD eff.}}},
\end{align*}
as well as contrasts between any the ICERs of any pair of regimes numbers $i,j \in \{2, \ldots ,15\}$, $i\neq j$:
\begin{align*}
     \psi_{\text{contrast}} = \psi_{\text{ICER}_{\tilde{d}^{(i)}}} - \psi_{\text{ICER}_{\tilde{d}^{(j)}}}.
\end{align*}

\section{Estimation and Inference}\label{estandinf}

\subsection{Estimation}\label{est}

In order to estimate the statistical estimands presented in the previous section, we consider two estimators that are consistent for the longitudinal treatment specific mean cost, i.e., $\psi_{\text{cost}, \tilde{d}}$: 1) an IPW estimator,
\begin{align*}
\hat{\psi}_{\text{IPW, cost}, \tilde{d}} & =\frac{1}{n}\sum_{i=1}^n\frac{\mathbb{I}[\bar{A}_i(K)=\tilde{\bar{d}}_K(\bar{Z}_i(K))]}{\prod_{t=1}^Kg_n(A_i(t)|\bar{X}_i(t), \bar{A}_i(t-1))}C_i,
\end{align*}
where the $g_n$ are estimates of the treatment mechanism factors $g_0$. Here, $g_n$ could be obtained by a maximum likelihood estimate based on a correctly specified lower-dimensional parametric model (such as a logistic regression with $\bar{Z}(t)$ alone, or with $\bar{Z}(t)$ plus additional covariates $\bar{X}(t)$) or the true treatment mechanism $g_0$, which is known in a SMART. The latter approach is often called the ``weight and replicate" method \citep{nahum2012experimental, almirall2014introduction}. A longitudinal TMLE based on iterated conditional expectations provides an alternative:
\begin{align*}
    \hat{\psi}_{\text{TMLE, cost}, \tilde{d}} = & \frac{1}{n}\sum_{i=1}^n\hat{\mathbb{E}}^*[\hat{\mathbb{E}}^*[ \ldots \\
    & \hat{\mathbb{E}}^*\left[\hat{\mathbb{E}}^*\left[C|\bar{X}(K)_i, \bar{A}(K) = \tilde{\bar{d}}_K(\bar{Z}(K))\right] \vert \bar{X}(K-1)_i, \bar{A}(K-1) = \tilde{\bar{d}}_{K-1}(\bar{Z}(K-1)) \right] \\
    & \ldots \vert X(1)_i, A(1) = \tilde{d}_1(Z(1)) ]],  
\end{align*}
which comprises iterated conditional expectations estimates that can be estimated using flexible machine learning algorithms and are then ``targeted" using the estimated or known treatment mechanisms in such a way that optimizes the bias-variance tradeoff for $\psi_{\text{cost}, \tilde{d}}$ \citep{bangrobins2005, van2012targeted}. We refer the reader to, for example, \cite{montoya2023efficient, tran2019double, schnitzer2013targeted} for step-by-step details on these longitudinal estimators; both can be implemented using the \emph{ltmle} package \citep{ltmlepackage}.

Once with $\hat{\psi}_{\text{IPW, cost}, \tilde{d}}$ or $\hat{\psi}_{\text{TMLE, cost}, \tilde{d}}$ in hand, one can estimate the risk difference between an active embedded regime and the SOC, the ICER of that embedded regime, and contrasts between the ICERs of two active embedded regimes. For example, using TMLE, the difference in expected monetary cost between an active regime $\tilde{d}$ and the SOC throughout $\tilde{d}_0$ is \begin{align*}
    \hat{\psi}_{\text{TMLE, RD cost}} = \hat{\psi}_{\text{TMLE, cost}, \tilde{d}} - \hat{\psi}_{\text{TMLE,cost}, \tilde{d}_0},
\end{align*} 
where $\hat{\psi}_{\text{TMLE, cost}, \tilde{d}_0}$ is the estimated expected cost under the embedded regime in which all receive SOC throughout (i.e., replacing $\tilde{d}$ with $\tilde{d}_0$ in the TMLE procedure described above). Then, replacing $C$ with $Y$ and multiplying by 100 in the above TMLE, the estimated risk difference for viral suppression is $\hat{\psi}_{\text{TMLE, RD eff.}} = \hat{\psi}_{\text{TMLE, eff.}, \tilde{d}} - \hat{\psi}_{\text{TMLE, eff.}, \tilde{d}_0}$. A TMLE estimate of the ICER is then:
\begin{align*}
     \hat{\psi}_{\text{TMLE, ICER}} = \frac{\hat{\psi}_{\text{TMLE, RD cost}}}{\hat{\psi}_{\text{TMLE, RD eff.}}},
\end{align*}
and contrast pairs for regimes $i,j \in \{2, \ldots ,15\}$, $i\neq j$ can be estimated via the following:
\begin{align*}
     \hat{\psi}_{\text{TMLE, contrast}} = \hat{\psi}_{\text{TMLE, ICER}_{\tilde{d}^{(i)}}} - \hat{\psi}_{\text{TMLE, ICER}_{\tilde{d}^{(j)}}}.
\end{align*}

\subsection{Inference}\label{inf}

Inference for the presented TMLE and IPW estimators can be based on the non-parametric bootstrap \citep{polsky1997confidence, briggs1997pulling} or the estimated efficient influence curve for the statistical parameters corresponding to the expected difference in embedded regimes, ICER, and contrasts between ICERs. Under assumptions, the variances of these efficient influence curves divided by $n$ are the variances of the limiting normal distribution of the aforementioned estimators. Given this, the empirical variances of the estimated efficient influence curves divided by $n$ can be used to construct Wald-type 95\% confidence intervals that provide nominal (if $g_0$ is used) to conservative (if $g_n$ is used) coverage for each parameter \citep{bangrobins2005, moore2009covariate}. We note that TMLE additionally has the potential for efficiency gains relative to IPW if its iterated conditional expectations are estimated consistently, which may occur if the iterated conditional expectations are estimated using machine learning (which TMLE allows) \citep{van2011targeted}. 

The efficient influence curves for $IC^*_{\text{cost}, \tilde{d}}$ corresponding to $\psi_{\text{cost}, \tilde{d}}$ and $IC^*_{\text{eff}, \tilde{d}}$ corresponding to $\psi_{\text{eff}, \tilde{d}}$ are presented in \cite{montoya2023efficient}. Then, the efficient influence curve for the expected cost between an active embedded regime and the standard of care $\psi_{\text{RD cost}}$ is simply $IC^*_{\text{RD cost}} = IC^*_{\text{cost}, \tilde{d}} - IC^*_{\text{cost}, \tilde{d}_0}$; the influence curve for the risk difference with respect to the clinical outcome, viral suppression, i.e., $\psi_{\text{RD eff.}}$, is $IC^*_{\text{RD eff.}} = IC^*_{\text{eff.}, \tilde{d}} - IC^*_{\text{eff.}, \tilde{d}_0}$. The true efficient influence curve for the ICER is: 
\begin{align*}
    IC^*_{\text{ICER}} &= \frac{1}{\psi_{\text{RD eff.}}}IC^*_{\text{RD cost}} - \frac{\psi_{\text{RD cost}}}{\psi_{\text{RD eff.}}^2}IC^*_{\text{RD eff.}},
\end{align*}
(the derivation of this can be found in Appendix A) and the influence curve for contrasts between the ICERs of two active regimes $i,j \in \{2, \ldots ,15\}$, $i\neq j$ is $IC^*_{\text{contrast}} = IC^*_{\text{ICER}_{\tilde{d}^{(i)}}} - IC^*_{\text{ICER}_{\tilde{d}^{(j)}}}$. 

The variance of the ICER estimator, as derived in \cite{o1994search}, is equivalent to the variance of $IC^*_{\text{ICER}}$ divided by $n$. The variance of the efficient influence curve for the ICER divided by $n$ can be written as follows \citep{chaudhary1996estimating}:
\begin{align*} 
Var(IC^*_{\text{ICER}})/n=
\psi_{\text{ICER}}^2\left(
\underbrace{\frac{Var(IC_{\text{RD cost}}^*)/n}{(\psi_{\text{RD cost}})^2}}_{\text{a}}
+ 
\underbrace{\frac{Var(IC^*_{\text{RD eff}})/n}{(\psi_{\text{RD eff}})^2}}_{\text{b}}
- 
2\frac{Cov(IC_{\text{RD cost}}^*,IC_{\text{RD eff}}^*)/n}{(\psi_{\text{RD eff}})(\psi_{\text{RD cost}})}
\right),
\end{align*}
where (a) is the square of the coefficient of variation of cost and (b) is the square of the coefficient of variation of the effect. \cite{briggs1997pulling} explain that these  coefficients of variation measure the relative proximity of the ICER's numerator or denominator (respectively) to zero: the higher the coefficient of variation, the closer the risk difference is to zero. This is of particular importance when evaluating the coefficient of variation for the effect; if it is large, that means the effect is closer to zero, which means the ICER will have a small denominator. In that case, estimates of the ICER will be unreliable, and estimator performance will be affected. Such performance dropoff will occur for both influence-curve based inference and non-parametric bootstrap-based inference \citep{briggs1997pulling}.

\section{Simulations}

Using simulations, we evaluated the performance of the IPW (using $g_0$, as in the weight and replicate method) and TMLE (using $g_n$) estimators described in Section \ref{est} for $\psi_{\text{ICER}}$, the true ICER. Inference was based on the efficient influence curve for both estimators, as described in Section \ref{inf}. For TMLE, the iterated conditional expectations were estimated using machine learning \citep{van2007super} and adjusted for all covariates, and $g$ was estimated using a correctly specified logistic regression model that included all baseline and time-varying covariates. We used the \emph{ltmle} R package for estimation and inference \citep{ltmlepackage, petersen2014causal}.  

We did this for a data generating process (DGP) corresponding to a SMART design with 8 embedded regimes (we used the first DGP presented in \cite{montoya2023efficient}); the first of these 8 regimes was defined as the SOC regime. Appendix B describes this DGP and the true parameter values, including the ICER values and individual values that comprise the ICER. Each simulation consisted of 500 iterations of $n$=1,809 observations (the complete sample size for ADAPT-R). We evaluated estimator performance in terms of bias, variance of estimates across simulation repetitions, average confidence interval width across simulation repetitions, and 95\% confidence interval coverage. Additionally, for every embedded regime, we calculated the estimated coefficient of variation for cost and effectiveness, and averaged each across simulation repetitions. For TMLE, we calculated the relative variance between TMLE and IPW by dividing the empirical variance of the TMLE estimates across simulation repetitions by the empirical variance of the IPW estimates across simulation repetitions. 

Simulation results are presented in detail in Table \ref{simresults}; several findings emerge from this study. First, for both IPW and TMLE, it is indeed the case that estimator performance suffers when the coefficients of variation are high. For example, embedded regimes 3, 5, and 7 have the highest average coefficients of variation, particularly corresponding to the effect. These regimes correspond to those that have the smallest effect compared to the SOC; as a consequence, performance suffers for estimators of these regimes. Importantly, these estimates are highly biased, suggesting that even if inference had been obtained with the non-parametric bootstrap, performance would still have suffered for estimators of those regimes. Second, among those regimes that do not have high coefficients of variation (i.e., 2, 4, 6, 8), both IPW and TMLE perform reasonably well (e.g., bias, variance, and MSE are low; confidence interval widths are stable and relatively small; and coverage is nominal), highlighting that the methods presented in this paper are valid for estimation and inference. However, among the regimes that yield acceptable estimator performance, IPW exhibits a higher variance than TMLE: the variance of the presented IPW estimator was 1.0008-1.0397 times that of TMLE's.

\begin{table}[h!]
    \centering
    \begin{subtable}[t]{\textwidth}
        \centering
\begin{tabular}{rrrrrrrr}
  \hline
 & 2 & 3 & 4 & 5 & 6 & 7 & 8 \\ 
  \hline
Bias & 0.0024 & -14.2228 & 0.0044 & -2.9353 & 0.0017 & -0.0317 & 0.0026 \\ 
  Var. & 0.0022 & 520.4834 & 0.0018 & 4125.6661 & 0.0020 & 67.5487 & 0.0017 \\ 
  MSE & 0.0022 & 722.7717 & 0.0018 & 4134.2820 & 0.0020 & 67.5497 & 0.0017 \\ 
  CI width & 0.1785 & 1883.7981 & 0.1689 & 92373.8118 & 0.1749 & 391.2452 & 0.1667 \\ 
  Cov. \% & 94.40 & 31.20 & 95.20 & 99.20 & 94.60 & 86.00 & 96.00 \\ 
  Avg. Coef. Var. (Cost) & 1.0607 & 0.2464 & 0.7895 & 11.2999 & 0.7096 & 1.8970 & 0.4724 \\ 
  Avg. Coef. Var. (Eff.) & 0.1116 & 21.4209 & 0.1194 & 366.9471 & 0.1040 & 12.2906 & 0.1110 \\ 
   \hline
\end{tabular}
        \caption{IPW simulation results.}
    \end{subtable}
    
    \vspace{1em} 

    \begin{subtable}[t]{\textwidth}
        \centering
\begin{tabular}{rrrrrrrr}
  \hline
 & 2 & 3 & 4 & 5 & 6 & 7 & 8 \\ 
  \hline
Bias & 0.0025 & -13.8582 & 0.0042 & -0.0047 & 0.0019 & 2.5262 & 0.0024 \\ 
  Var. & 0.0021 & 1498.8456 & 0.0017 & 1.2477 & 0.0020 & 2035.9392 & 0.0016 \\ 
  MSE & 0.0021 & 1690.8944 & 0.0018 & 1.2477 & 0.0020 & 2042.3210 & 0.0017 \\ 
  CI width & 0.1774 & 5664.5179 & 0.1669 & 18.7488 & 0.1743 & 4827.9197 & 0.1652 \\ 
  Cov. \% & 94.80 & 32.40 & 95.40 & 99.20 & 94.00 & 87.00 & 95.20 \\ 
  Avg. Coef. Var. (Cost) & 0.7096 & 0.2462 & 0.6416 & 16.3390 & 0.8023 & 0.7949 & 0.5021 \\ 
  Avg. Coef. Var. (Eff.) & 0.1057 & 38.1652 & 0.1127 & 4.6804 & 0.0984 & 27.7300 & 0.1047 \\ 
  Rel. Var. IPW & 1.0205 & 0.3473 & 1.0397 & 3306.6296 & 1.0008 & 0.0332 & 1.0362 \\ 
   \hline
\end{tabular}
        \caption{TMLE simulation results.}
    \end{subtable}

    \caption{Simulation results for performance of an inverse probability of treatment weighting (IPW; top table (a)) estimator and targeted maximum likelihood estimator (TMLE; bottom table (b)) for the incremental cost effectiveness ratio (ICER). Column names are active embedded regimes whose ICERs are with respect to the first embedded regime, the care standard. Metrics include bias, variance (Var.), mean squared error (MSE), confidence interval (CI) width, confidence interval coverage (Cov. \%), an average of the coefficients of variation for cost (Avg. Coef. Var. (Cost)), and an average of the coefficients of variation for the effect (Avg. Coef. Var. (Eff.)). In the bottom table (b), we additionally present the relative variance between TMLE and IPW.}
    \label{simresults}
\end{table}

\section{Results}

The goal of this analysis was to understand the cost and cost effectiveness of the regimes embedded within ADAPT-R. We carried out this analysis using the TMLE presented above for estimation with influence curve-based inference (confidence intervals) throughout. 

\begin{figure}[h]
    \centering
    \includegraphics[scale = .75]{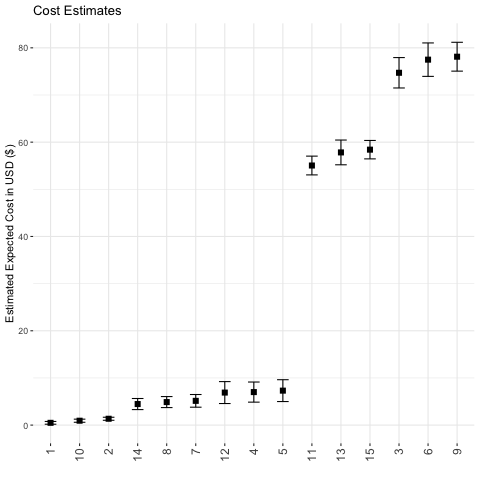}
    \caption{Targeted maximum likelihood estimates of the expected cost in USD (\$) of each embedded dynamic treatment regime within the Adaptive Strategies for Preventing and Treating Lapses of Retention in HIV Care (ADAPT-R) trial in order of least to most costly. Squares denote point estimates and error bars are 95\% confidence intervals. The $x$-axis depicts the embedded regime number corresponding to the list in Table \ref{table0}.}
    \label{cost}
\end{figure}

First, we estimated the expected cost of each of the embedded regimes (Figure \ref{cost}), i.e., $\hat{\psi}_{\text{TMLE, cost, d}}$. The regime with the highest estimated expected cost was regime \#6: CCTs, augmented by SMS messages if a patient has a lapse in care, otherwise continue CCTs (estimate: \$78.11, CI: \$75.06, \$81.18). Indeed, this embedded regime was significantly more expensive than the standard of care throughout (difference: \$77.01, CI: \$73.45,\$80.56). The least expensive embedded regime (after the care standard throughout) was \#10, which administers SMS messages to all initially, followed by SOC outreach if a patient has a lapse in care, and discontinuation of SMS messages if there is no lapse in care (estimate: \$0.93, CI: \$0.60, \$1.26). This regime was not found to be significantly more expensive than the SOC throughout (difference: \$0.45, CI: \$0.01,\$0.89).

In Table \ref{iceradapt}, we show all ICER estimates (including their numerator and denominator risk difference estimates) with confidence intervals for each embedded regime. In this table, we additionally present the coefficients of variation for each embedded regime. Embedded regimes 12 (SMS, followed by SMS and CCTs if lapse; otherwise stop SMS), 13 (CCTs, followed by SMS and CCTs if lapse; otherwise stop CCTs), and 15 (CCTs, followed by navigator if lapse; otherwise stop CCTs) had the highest coefficients of variation. Per suggestions in the literature (e.g., \cite{chaudhary1996estimating, briggs1997pulling}) and our simulation study, an estimator is unlikely to be reliable if its coefficients of variation are high. Thus, from here forward, we will only discuss results for embedded regimes that had coefficients of variation $< 2$.

\begin{table}[ht]
\centering
\begin{tabular}{llrrrr}
  \hline
Regime \# & $\hat{\psi}_{\text{TMLE, ICER}}$ [95\% CI] & $\hat{\psi}_{\text{TMLE, RD cost}}$ & $\hat{\psi}_{\text{TMLE, RD eff}}$ & Coef. Var. (Cost) & Coef. Var. (Eff.) \\ 
  \hline
2 & 0.23 [-0.20, 0.66] & 0.88 & 3.79 & 0.26 & 0.89 \\ 
  3 & 12.63 [0.07, 25.19] & 76.05 & 6.02 & 0.02 & 0.51 \\ 
  4 & 1.36 [0.32, 2.39] & 7.48 & 5.52 & 0.17 & 0.44 \\ 
  5 & 1.73 [-1.62, 5.07] & 5.72 & 3.31 & 0.17 & 0.98 \\ 
  6 & 11.74 [1.29, 22.19] & 79.46 & 6.77 & 0.02 & 0.46 \\ 
  7 & 0.95 [0.00, 1.90] & 4.50 & 4.73 & 0.15 & 0.49 \\ 
  8 & 0.59 [0.10, 1.08] & 4.92 & 8.29 & 0.14 & 0.38 \\ 
  9 & 9.21 [2.92, 15.49] & 79.16 & 8.60 & 0.02 & 0.35 \\ 
  10 & 0.23 [-0.59, 1.05] & 0.45 & 1.96 & 0.50 & 1.71 \\ 
  11 & -31.88 [-146.22, 82.46] & 55.63 & -1.74 & 0.02 & 1.83 \\ 
  12 & 3.98 [-14.68, 22.64] & 5.27 & 1.33 & 0.19 & 2.40 \\ 
  13 & -61.72 [-470.33, 346.89] & 59.00 & -0.96 & 0.02 & 3.37 \\ 
  14 & 0.71 [-0.01, 1.42] & 4.60 & 6.52 & 0.16 & 0.47 \\ 
  15 & 72.56 [-478.18, 623.30] & 58.67 & 0.81 & 0.02 & 3.87 \\ 
   \hline
\end{tabular}
\caption{Incremental cost effectiveness ratio (ICER) analysis results for the active embedded dynamic treatment regimes within the Adaptive Strategies for Preventing and Treating Lapses of Retention in HIV Care (ADAPT-R) trial. The regime numbers (first column) correspond to those in Table \ref{table0}. Point estimates of the ICERs estimated with targeted maximum likelihood estimation (TMLE) and 95\% confidence intervals (in brackets) are presented in the second column; the unit for each is USD (\$) per per additional person with viral suppression. The numerators (cost risk difference [RD]; units are USD [\$]) and denominators (effect RD) of these ICERs are presented in the subsequent two columns. Coefficients of variation for cost (Coef. Var. [Cost]) and effect (Coef. Var. [Eff.]) are presented in the last two columns.}
    \label{iceradapt}
\end{table}

\begin{figure}[h]
    \centering
    \includegraphics[scale = .75]{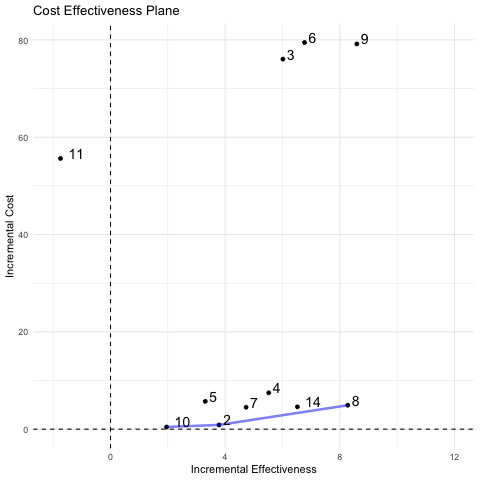}
    \caption{Cost effectiveness plane for the dynamic treatment regimes embedded in the Adaptive Strategies for Preventing and Treating Lapses of Retention in HIV Care (ADAPT-R) trial. Each point corresponds to an incremental cost effectiveness ratio that is estimated using targeted maximum likelihood estimation. The numbers next to each point correspond to the embedded regimes listed in Table \ref{table0}. The blue line depicts the efficient frontier \citep{suen2016efficient}.}
    \label{quads}
\end{figure}

For the remaining regimes, we present a plot of the incremental cost versus the incremental effectiveness (Figure \ref{quads}). Not accounting for statistical uncertainty, all estimates appear in the first quadrant of the graph, meaning that all regimes were both more effective and expensive than the SOC, except for regime \#11, which gives CCTs initially, followed by SOC outreach if there is a lapse in care, otherwise stop CCTs, which appears to be less effective than SOC, but still more costly. Results show that the smallest ICER estimate in the first quadrant (lowest non-negative additional cost per additional person achieving viral suppression, interpreted as the most cost effective embedded regime among the effective and costly interventions versus SOC) corresponds to regime \#10, which gives all patients SMS, followed by SOC outreach if there is a lapse in care, otherwise discontinue SMS messages (ICER = \$0.23 per additional person with viral suppression, CI: -\$0.59, \$1.05). In contrast, the embedded regime with the highest ICER (interpreted as the least cost effective embedded regime among the effective and costly interventions versus SOC), regime \#3, says: give everyone CCTs, replace the CCT with SOC outreach if there is a lapse, otherwise continue CCTs (ICER = \$12.64 per additional person with viral suppression, CI: \$0.06, \$25.23).

The ICERs corresponding to the top 3 most effective embedded regimes within ADAPT-R were 1) \$9.21 per additional person with viral suppression (CI: \$2.91, \$15.50) corresponding to regime \#9 (CCTs, followed by a peer navigator if there is a lapse, otherwise continue CCTs); 2) \$0.59 per additional person with viral suppression (CI: \$0.10, \$1.09) corresponding to regime \# 8 (SMS, followed by a peer navigator if there is a lapse, otherwise continue SMS); and 3) \$11.74 per additional person with viral suppression (CI: \$1.28, \$22.19) corresponding to regime \#6 (CCT, followed by SMS and CCTs if there is a lapse, otherwise continue CCTs). 

Lower ICERs in the first quadrant correspond to more cost effectiveness; two of the aforementioned top 3 most \emph{effective} embedded regimes within ADAPT-R were significantly less cost effective than the most \emph{cost effective} embedded regime (i.e., regime \#10: SMS followed by SOC outreach if lapse; discontinue SMS otherwise) in ADAPT-R. These included a) the most effective strategy, regime \#9: CCT followed by peer navigator if lapse; otherwise continue (ICER difference: \$8.98 per additional person with viral suppression, CI: \$3.03, \$14.93) and b) the third-most effective strategy, regime \#6: CCT followed by CCT+SMS if lapse; otherwise continue CCT (ICER difference: \$11.51 per additional person with viral suppression, CI: \$1.40, \$21.62). There was insufficient evidence to conclude that the second-most effective strategy (regime \#8: SMS followed by peer navigator if lapse; otherwise continue SMS) was more or less cost effective than the most cost effective regime (ICER difference: \$0.37 per additional person with viral suppression, CI: -\$0.33, \$1.06). 

\section{Discussion}

In this paper, we detailed an estimation and inference procedures for cost effectiveness analyses using data from a SMART design. In particular, we presented a semiparametric efficient estimator (i.e., TMLE; \cite{van2006targeted}) for the ICER of a SMART's embedded regimes, offering an approach that has the potential to improve precision compared to the methods commonly proposed in cost effectiveness analyses for SMARTs. We described how inference for this TMLE can be obtained using an influence curve approach, which can be more computationally feasible than the often-proposed non-parametric bootstrap. A simulation study highlighted both the potential utility of using these methods for this type of analysis and scenarios under which caution should be taken when estimating ICERs (i.e., when effects are small). 

Throughout, we used as illustration a cost effectiveness analysis for the ADAPT-R trial, presenting estimated ICERs (with inference) for embedded regimes aimed at increasing HIV care adherence. While the most \emph{effective} strategies in ADAPT-R require adapting to patients' care status by maintaining an active first-line treatment if there is no lapse or augmenting/replacing it with another active treatment if there is a lapse, the most \emph{cost effective} strategy only required a relatively inexpensive first-line treatment (SMS) without an active second-line treatment. At the same time, the second-most effective treatment, which requires an active treatment throughout (initiate with SMS; replace with navigator if lapse, otherwise continue SMS) was not significantly different than the most cost effective strategy -- implying the possibility of obtaining maximal cost effectiveness with one of the most effective embedded strategies. 

We note that many in the literature have discussed challenges in obtaining inference for ICER estimates \citep{chaudhary1996estimating, heitjan1999problems, briggs1997pulling, wang2008study, fan2007simulation}. In particular, in this paper we follow the so-called ``Taylor series" approach to obtaining inference by relying on the assumption that the sampling distribution of both the presented IPW and TMLE will converge to a normal distribution, allowing for Wald-type 95\% confidence intervals. When effects are close to zero, the denominator of the ICER is small, affecting the ICER estimator and thus its limiting distribution \citep{chaudhary1996estimating}. Some have proposed the non-parametric bootstrap (and variations of it) as an alternative; however, bootstrap-type methods face issues when the ICER's denominator is small, as well \citep{heitjan1999problems, briggs1997pulling}. As demonstrated in this paper, calculating coefficients of variation based on influence curve estimates can offer a diagnostic for the validity of the presented approaches.

SMARTs hold appeal because they allow for the evaluation of adaptive strategies that have to potential to be, in theory, more cost effective than ``one-size-fits-all" strategies (i.e., strategies that give all persons the same treatment regardless of their measured history, such as response to an initial treatment). This is because such strategies more selectively administer treatment to only those who need it, thereby saving resources by not administering treatment among those who do not need it. The methods presented in this paper offer a way to quantify this, moving us closer to understanding the most effective and efficient strategies to improve public health outcomes.

\section*{Acknowledgments}
Research reported in this publication was supported by NIH awards R01AI074345 and R00MH133985. The content is solely the responsibility of the authors and does not necessarily represent the official views of the NIH. We thank Shalika Gupta for helpful discussions during the research stage of this work.

\bibliographystyle{plainnat}
\bibliography{wileyNJD-AMA}

\appendix

\section{Efficient influence curve for the ICER}

As described in \citep{TLBBD}, for a collection of parameters $\psi_j$ with respective influence curves $IC_j$ for $j = 1,\ldots,J$, one can obtain the influence curve of a function of this parameter collection $f(\psi_j:j)$ as follows: $\sum_j \frac{\partial}{\partial \psi_j} f(\psi_j:j) IC_j$. Our parameter of interest is a function (ratio) of two risk differences: $f(\psi_{\text{RD cost}},\psi_{\text{RD eff.}}) = \frac{\psi_{\text{RD cost}}}{\psi_{\text{RD eff.}}}$. Then:

\begin{align*}
    IC^*_{\text{ICER}} &= \frac{\partial}{\partial \psi_{\text{RD cost}}} f(\psi_{\text{RD cost}},\psi_{\text{RD eff.}}) IC^*_{\text{RD cost}} + \frac{\partial}{\partial \psi_{\text{RD eff.}}} f(\psi_{\text{RD cost}},\psi_{\text{RD eff.}}) IC^*_{\text{RD eff.}} \\
     &= \frac{\partial}{\partial \psi_{\text{RD cost}}} \frac{\psi_{\text{RD cost}}}{\psi_{\text{RD eff.}}} IC^*_{\text{RD cost}} + \frac{\partial}{\partial \psi_{\text{RD eff.}}} \frac{\psi_{\text{RD cost}}}{\psi_{\text{RD eff.}}} IC^*_{\text{RD eff.}} \\
     &= \frac{1}{\psi_{\text{RD eff.}}}IC^*_{\text{RD cost}} - \frac{\psi_{\text{RD cost}}}{\psi_{\text{RD eff.}}^2}IC^*_{\text{RD eff.}},
\end{align*}
which is the conjectured efficient influence curve of the ICER.

\section{Data generating process for simulations}

Data generating process (DGP) for a simple SMART with 8 embedded regimes in which re-randomization is based only on intermediate covariates, the covariates, treatments and outcome:

\begin{align*}
    X(1) &\sim Normal(\mu = 0,\sigma = 1) \\
    A(1) & = Bernoulli(p = 0.5) \text{ with support \{0,1\}}\\
    L(2) &\sim Bernoulli(p = \textrm{expit}(X(1) + A(1)) \\
    S(2) &\sim Normal(\mu = X(1) + 2A(1), \sigma = 1) \\
    A(2) & \sim \begin{cases}
        \text{If } L(2) = 1,   Bern(p = 0.5) \text{ with support \{1,2\}}\\ 
        \text{If } L(2) = 0,   Bern(p = 0.5) \text{ with support \{3,4\}}\\
        \end{cases} \\
    Y & \sim Bernoulli(p = \logit^{-1}(\logit(y) + S(2) + 0.5X(1)^2 + \log(|X1|+.01))))\\
    C & \sim 5Exp(\lambda = c + |S(2) + X(1) + L(2) - 3*A(1))|
\end{align*}
where $y = 1 - (.28, .26, .28, .3, .29, .3, .21, .2)$ (for the outcome $Y$) and $c = (2, 0.03, 0.035, 0.044, 0.06, 0.05, 0.058, 0.025)$ (for cost $C$). These are vectors of fixed constants unique to each of the 8 embedded regimes (respectively). Here, $(L(2), S(2)) = X(2)$. The true values of each of the causal parameters described in the main text (Section \ref{causalparams}) are presented in Table \ref{tableDGP1truevals}.

\begin{table}[ht]
\centering 
\begin{tabular}{lrrrrr}
  \hline
Regime \# & $\mathbb{E}_{P_{U,X}}[Y_{\tilde{d}}]$ & $\mathbb{E}_{P_{U,X}}[C_{\tilde{d}}]$ & $\psi_{\text{RD cost}}$ & $\psi_{\text{RD eff}}$ & $\psi_{\text{ICER}}$ \\ 
  \hline
1 (SOC) & 0.6050 & 3.9686 & -- & -- &  -- \\ 
  2 & 0.8637 & 7.0779 & 3.1094 & 25.8660 & 0.1202 \\ 
  3 & 0.6067 & 6.2592 & 2.2906 & 0.1650 & 13.8825 \\ 
  4 & 0.8517 & 6.6183 & 2.6497 & 24.6610 & 0.1074 \\ 
  5 & 0.6392 & 4.0193 & 0.0508 & 3.4140 & 0.0149 \\ 
  6 & 0.8771 & 7.2908 & 3.3223 & 27.2090 & 0.1221 \\ 
  7 & 0.6424 & 6.3026 & 2.3341 & 3.7340 & 0.6251 \\ 
  8 & 0.8646 & 6.8548 & 2.8863 & 25.9580 & 0.1112 \\ 
   \hline
\end{tabular}
\caption{True causal parameter values of each of the embedded regimes in the simulated DGP.}
\label{tableDGP1truevals}
\end{table}




\end{document}